\documentclass[preprint]{vgtc}               

\graphicspath{{figures/}{pictures/}{images/}{./}} 

\usepackage{times}                     

\usepackage{tabu}                      
\usepackage{booktabs}                  
\usepackage{lipsum}                    
\usepackage{mwe}                       
\usepackage{amsmath}
\usepackage{amssymb}
\usepackage{multirow}
\usepackage{algorithm}
\usepackage{algorithmic}
\usepackage{adjustbox}
\usepackage{mathptmx}                  

\onlineid{0}

\vgtccategory{Research}

\vgtcinsertpkg

\title{Extend Your Horizon: A Device-Agnostic Surgical Tool Tracking Framework with Multi-View Optimization for Augmented Reality}

\author{Jiaming Zhang \thanks{jzhan282@jh.edu; corresponding author}\\ %
	\scriptsize Johns Hopkins \\ \scriptsize University \\ %
	\and Mingxu Liu \\ %
	\scriptsize Johns Hopkins \\ \scriptsize University \\ %
	\and Hongchao Shu \\ %
	\scriptsize Johns Hopkins \\ \scriptsize University \\ %
    \and Ruixing Liang \\
    \scriptsize Johns Hopkins \\ \scriptsize University \\ %
	\and Yihao Liu \\ %
	\scriptsize Johns Hopkins \\ \scriptsize University \\ %
	\and Ojas Taskar \\ %
	\scriptsize Johns Hopkins \\ \scriptsize University \\ %
	\and Amir Kheradmand \\ %
	\scriptsize Johns Hopkins University, \\ \scriptsize School of Medicine %
	\and Mehran Armand \\ %
	\scriptsize University of Arkansas, \\ \scriptsize Johns Hopkins University %
	\and Alejandro Martin-Gomez \thanks{alejandro.martin@uark.edu; senior author}\\ %
	\scriptsize University of Arkansas, \\ \scriptsize Johns Hopkins University}

\teaser{
  \centering
  \includegraphics[width=\linewidth]{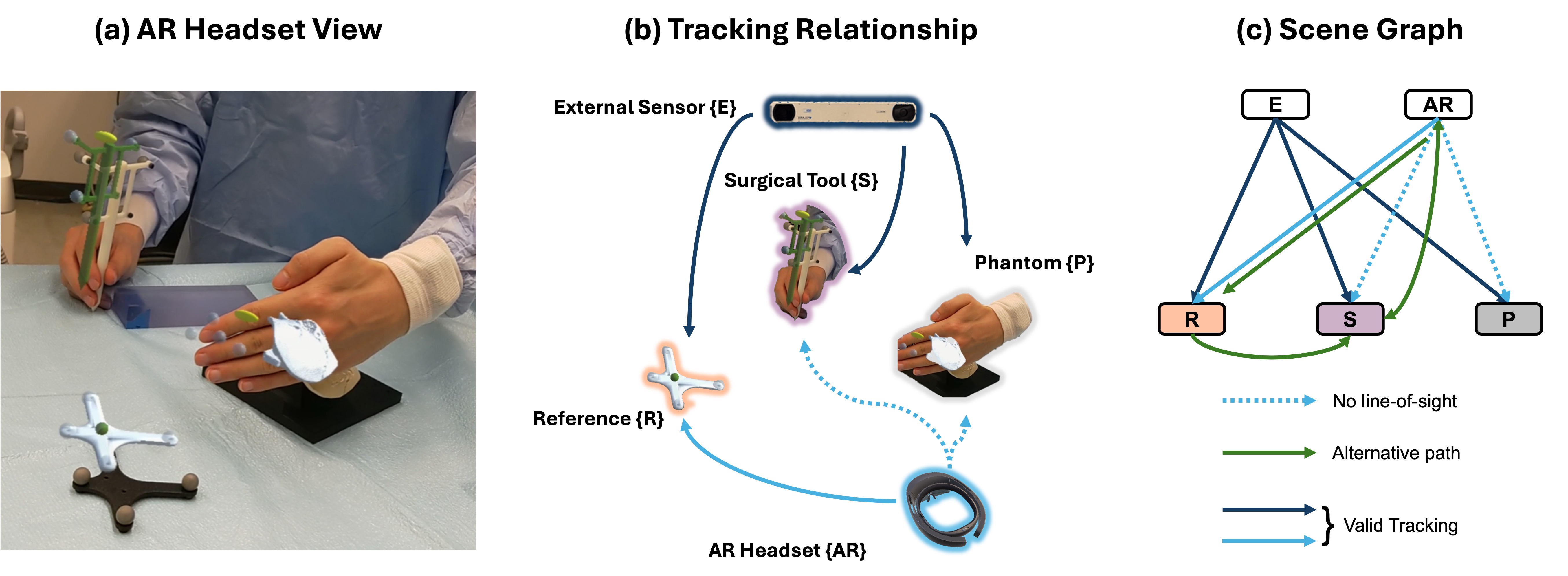}
  \caption{Overview of the proposed multi-sensor, multi-object tracking framework for surgical navigation. (a) User-perspective view from the Microsoft HoloLens 2 illustrating a challenging scenario in which two objects (or markers) are no longer directly tracked due to occlusion or movement beyond the sensor’s effective range. When direct tracking is lost, the proposed system preserves continuous object rendering by exploiting alternative kinematic relationships, with pose uncertainty visualized as yellow ellipsoids; when objects are directly tracked, their poses are displayed as green spheres. (b) Tracking status of three objects, namely the reference frame \{R\}, surgical tool \{S\}, and phantom \{P\}, as jointly observed by an AR headset and an external tracking system. (c) Dynamic scene graph representation of the same tracking scenario, illustrating how the pose of an occluded object is inferred by identifying an alternative path within the graph. The apparent misalignment between physical objects and their virtual overlays results from the screenshot being captured from an external camera viewpoint, which is offset from the user’s egocentric perspective; from the user’s viewpoint, the virtual overlays coincide with the physical objects.}
  \label{fig:teaser}
}

\abstract{
   Surgical navigation has proven to be an effective approach for providing real-time guidance and visualization of relevant information by estimating the pose of the patient’s anatomy and surgical tools. During navigated surgery, instruments are commonly equipped with fiducial markers and tracked by stationary optical tracking systems (OTS) to provide accurate navigation cues. Augmented Reality (AR) has been adopted for intuitive visual guidance, even motivating several efforts to enable surgical instrument tracking through built-in sensors on Head-Mounted Displays (HMDs). However, existing tracking methods typically require a direct line-of-sight to instruments, which is challenging to maintain in dynamic surgical environments due to frequent occlusions caused by moving medical equipment, surgical tools, and personnel. To address this challenge, this work introduces a novel framework capable of tracking surgical instruments even under occlusion by fusing different sensors in a dynamic scene graph representation. Our framework uniquely combines tracking systems with varying degrees of accuracy, providing real-time assessments of tracking reliability to the user. Unlike conventional sensor fusion approaches that are heavily dependent on specific sensor modalities, the proposed method is agnostic to tracking device modality and robust to their motion states (e.g., stationary OTS versus dynamic AR-HMD). Experimental results demonstrate that our dynamic scene graph framework successfully integrates and optimizes measurements from multiple tracking sources, significantly enhancing AR visualization consistency and accuracy with robustness under occlusion.
} 

\keywords{Mixed / Augmented Reality, Tracking, Dynamic graph algorithms}

\begin{document}

\firstsection{Introduction}
\maketitle

Augmented Reality (AR) has recently become a valuable tool in surgical navigation, providing surgeons with a continuous, in-situ visualization of critical structures and tool positions, effectively merging medical images with the live surgical view. AR systems require precise alignment between the virtual and real worlds to deliver consistent information and a seamless visual experience \cite{uchiyama2012}. This alignment can be achieved with sensors such as monocular cameras that capture the position and orientation of real-world objects \cite{rabbi2016}. However, single-camera approaches often fail under occlusion, reference drift, or appearance changes, which are commonly seen in dynamic environments \cite{lee2017}. The dependence on camera viewpoint \cite{dockstader2001} and the need for direct line-of-sight (LoS) between target objects and sensors further exacerbate these limitations \cite{zhang2022b}.

To mitigate occlusion, three main strategies have been explored. Model-based tracking approaches, such as SORT \cite{bewley_simple_2016}, employ Kalman filtering under the assumption of constant velocity, with adaptive variants refining predictions by estimating noise statistics online \cite{li_adaptive_2024}. Existing studies have encoded temporal dynamics directly from data, enabling more complex motion modeling. For example, Re$^3$ \cite{gordon2018re} integrates a Long Short-Term Memory to maintain robust predictions through occlusions \cite{milan_online_2017, fang_recurrent_2018}. In parallel, occlusion-aware networks are used to explicitly detect and adapt to partial or full occlusions, contributing to enhanced tracking reliability \cite{gupta_tackling_2021, zhuDistractorAware2018}.

Despite these continued advances, surgical environments pose unique challenges that expose deployment-level limitations. Optical tracking systems (OTS), which are widely adopted in clinical practice, are inherently susceptible to marker occlusion ( caused by surgical instruments, personnel, or medical devices), leading to intermittent tracking failure and workflow disruption \cite{mehbodniya_frequency_2019}. To mitigate occlusion, previous approaches have used multiple calibrated stationary sensors to provide redundant coverage \cite{ren2014}. However, while effective in controlled settings, such approaches rely on precise extrinsic calibration across a limited set of sensor types. In practice, even minor sensor displacement or intraoperative repositioning can invalidate this calibration and compromise the entire tracking pipeline, making these approaches less suitable for surgical AR scenarios involving mobile devices such as AR-HMDs.

In this work, we aim to address a new problem: tracking objects with non-stationary sensors, without explicitly knowing their spatial relationship. Compared with the existing systems, which are typically designed to expect relatively static sensor layouts, we introduce a multi-sensor and device-agnostic tracking framework suitable for AR applications. The framework represents sensors and tracked targets as nodes in a two-layer dynamic graph and formulates tracking as querying edges between these nodes. This abstraction decouples tracking continuity from fixed sensor configurations and optimizes spatial relationships in real time. As a result, the framework remains resilient to sensor motion, partial or complete occlusion, and temporary sensor failure. To improve generalizability, the proposed framework provides a generic interface to adapt existing OTS systems, AR devices, and other sensors. Our framework ensures robust visualization of tracked objects and informs users of the tracking uncertainty, ultimately supporting reliable AR integration in complex environments, including those observed during surgical procedures.

\begin{figure*}[htbp]
  \centering 
  \includegraphics[width=0.9\textwidth]{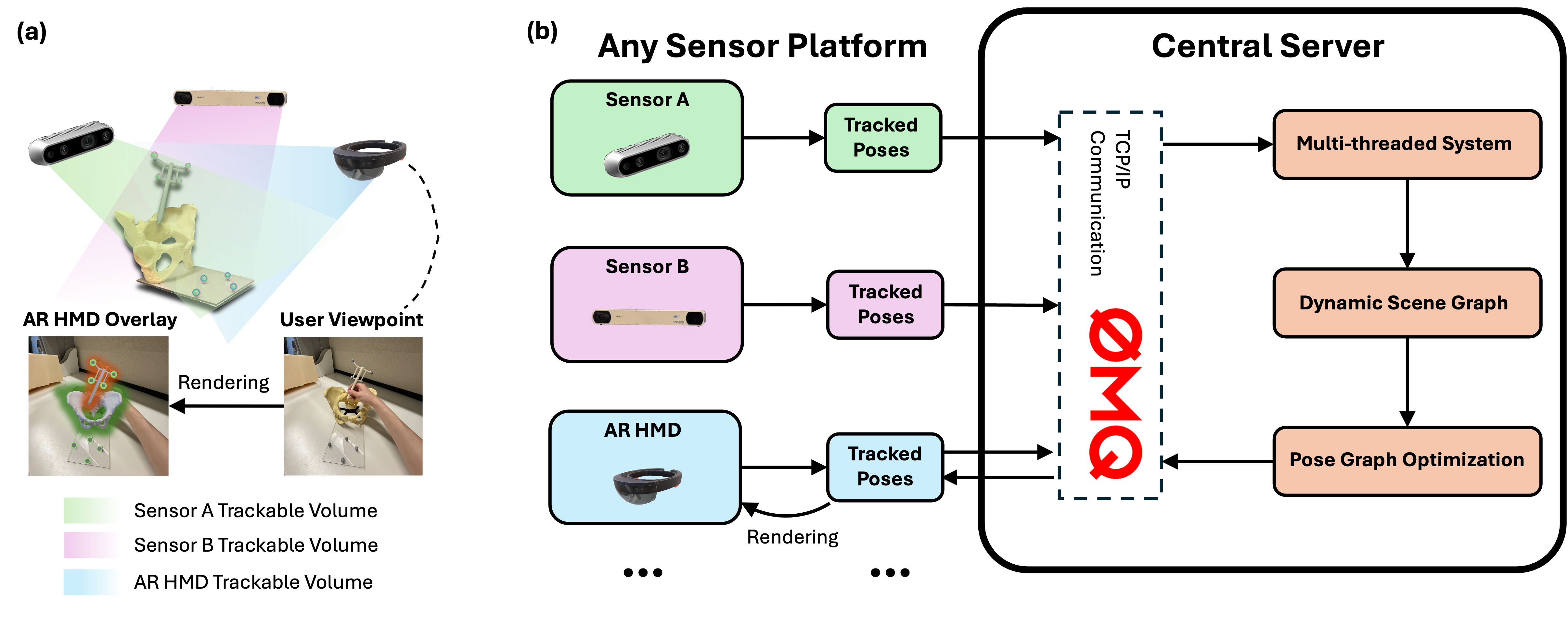}
  \caption{%
We integrate numerous trackers of diverse types into a standard surgical tracking scenario, ensuring continuity during clinical procedures and preventing track loss. (a) In cases of occlusion, our framework autonomously searches plausible paths within the Dynamic Scene Graph to reestablish tracking continuity.  When direct tracking is unattainable, our framework delivers fusion uncertainty estimations, enabling the completion of the AR scene for surgical guidance assistance. (b) The architecture of our framework has a central server that communicates with other sensors through TCP/IP. Each sensor tracks surgical tools independently and updates its tracking results based on the feedback from the central server. AR HMD works as a sensor and a visualization tool to render the tracked tools with their optimized poses.
  }
  \label{fig:Overview}
\end{figure*}

\section{Related Work}
In computer-aided surgery, AR has been widely adopted to improve situational awareness by enriching the surgical scene with digital content directly within the surgeon’s field of view. The use of Head-Mounted Displays (HMDs), in particular, provides advanced visualization capabilities by overlaying pre- and intraoperative information onto the operative field, enabling surgeons to perceive the relative positions of surgical instruments and anatomy more intuitively \cite{jung2022, chiou2022, muller_augmented_2020, jeung2023}. These capabilities have been shown to enhance intraoperative decision-making and reduce cognitive load.

When combined with tracking technologies, AR-HMDs further support accurate navigation. Surgical instruments are often equipped with retro-reflective fiducial markers tracked by OTS, which precisely estimate the position and orientation of instruments in real time, thereby assisting surgeons in achieving safer and more accurate procedures \cite{martin-gomez2023}. Beyond OTS-based tracking, AR-enabled systems have demonstrated accuracy comparable to OTS in controlled studies \cite{chiou2022, gsaxner2021, black2023, martin-gomez2023}, with variations in workflow suitability depending on the surgical context \cite{ma2023, dho2024}. However, these tracking methods that rely on a single device are inherently constrained by a limited tracking volume; their robustness deteriorates outside of this range, and they are highly susceptible to occlusions in dynamic surgical environments \cite{mehbodniya_frequency_2019}.

To address these issues, hybrid methods that integrate multiple sensing technologies have been proposed. Multi-camera fusion, for example, improves robustness by compensating for occlusions \cite{wang2019}, while combining stereo vision with Extended Kalman Filters (EKF) has shown utility in surgical navigation \cite{gsaxner2021}. In robot-assisted procedures, odometry fused with stereo vision and arthroscopic imagery has enabled continuous tracking with scale recovery \cite{marmol_dense-arthroslam_2019}. More general sensor fusion techniques, such as graph-based optimization, align multi-sensor information by minimizing global pose errors \cite{tao_multi-sensor_2022}. Although these approaches have achieved success within their specific use cases, they are typically designed for narrowly defined scenarios and lack generalizability across diverse surgical settings.

\begin{figure}[ht]
    \centering 
    \includegraphics[width=0.95\columnwidth]{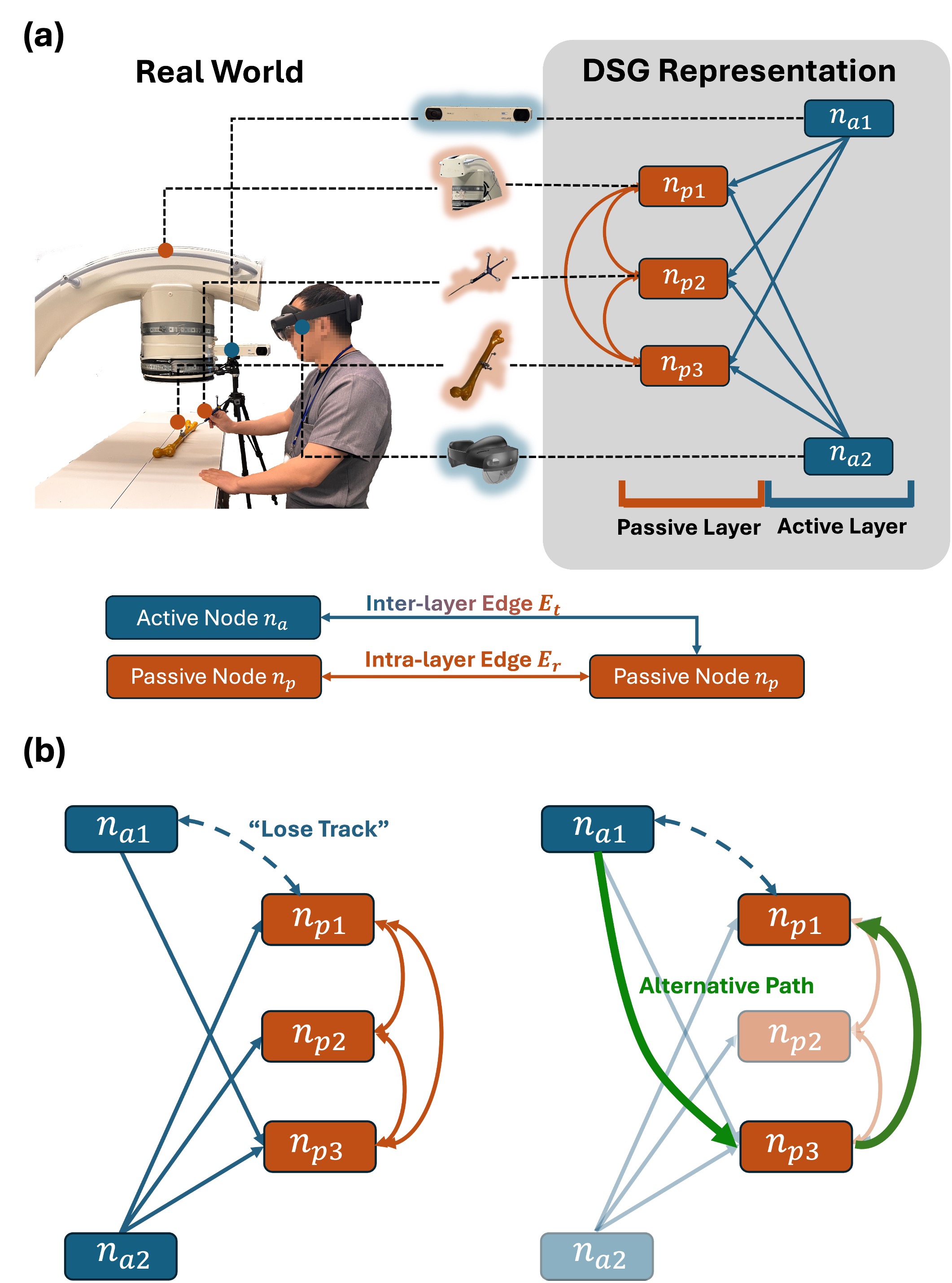}
        \caption{ (a) A mapping from a mock AR-guided orthopedic surgery to DSG representation. The surgeon with an AR-HMD interacts with a bone phantom positioned on an X-ray table under a C-arm device. Both the phantom and the surgeon's pointer are tagged with retro-reflective markers to enable tracking by OTS and AR-HMD. Within the DSG, passive nodes such as the C-arm, bone model, and pointer are depicted using $ n_{pj}, j \in {1,2,3} $ (highlighted in orange), while active nodes like the OTS and AR-HMD are denoted as $ n_{ai}, i \in {1,2}$ (in dark blue). The inter-layer edges bridge the $n_{ai}$ and $n_{pj}$, meaning the passive nodes are being tracked by the active nodes. The intra-layer edges connect only $n_{pj}$, representing the relative pose between the passive nodes. (b) demonstrates that when a direct tracking from $n_{a1}$ to $n_{p1}$ is lost, we can still track it indirectly through querying a path within the DSG.}
\label{fig:DSG}
\end{figure}

\section{Methods}

In this work, we propose a novel framework (\cref{fig:Overview}) that unifies multiple trackers and tracked entities into a Dynamic Scene Graph (DSG), which is subsequently optimized using Pose Graph Optimization (PGO) \cite{carlone_angular_2014}. Note that PGO operates over the node poses embedded in the graph rather than altering the graph topology itself. The framework enables continuous visualization of occluded entities by leveraging alternative kinematic chains and is inherently device-agnostic, allowing seamless integration of heterogeneous tracking devices. In addition, a universal TCP/IP interface ensures compatibility across modalities, addressing generalizability issues and enhancing adaptability in diverse environments.  

\subsection{Dynamic Scene Graph Representation}

Scene graphs are hierarchically structured directed graphs in which nodes represent entities and edges capture relationships among them \cite{rosinol2020}. Widely adopted in AR for context awareness \cite{tahara_retargetable_2020} and real-time rendering \cite{kalkofen_comprehensible_2009}, they naturally extend to DSGs that represent evolving scenes with moving entities such as surgical tools or patient anatomy.  
For marker-based tracking, we define a \textbf{two-layer DSG} (illustrated in \cref{fig:DSG}):  
\begin{itemize}
    \item \textbf{Active layer:} contains nodes ($n_{ai}$) representing trackers/sensors (e.g., OTS, RGB-D cameras, AR-HMDs).  
    \item \textbf{Passive layer:} contains nodes ($n_{pj}$) representing tracked entities (e.g., surgical tools, anatomy, reference markers).  
\end{itemize}

\noindent
The DSG construction follows three assumptions:  (1) The set of passive entities in the scene is finite and predefined. (2) Only active entities can directly localize passive entities; passive nodes cannot track each other.  (3) Active entities have overlapping fields of view (each device tracks independently).

\subsection{Edge Definitions}

Relationships between nodes are categorized into two types:  

\textbf{Inter-layer edges} represent direct measurements from a tracker to an object (active $\rightarrow$ passive). Each edge stores (1) pose in $\mathbf{SE}(3)$, (2) timestamp, and (3) binary tracking status. For each tracker--object pair, only one edge exists (no parallel edges).  

\textbf{Intra-layer edges}  represent relative transformations between passive entities (passive $\leftrightarrow$ passive). Since passive nodes cannot track each other, these edges are inferred by querying their measurements against a common tracker. Intra-layer edges may exist in parallel, allowing multiple trackers to estimate the same relative pose. The redundancy naturally models measurement noise, which can later be optimized via PGO.  

Unlike conventional systems that require a fixed world frame, the proposed DSG adopts a decentralized representation. Each passive node’s position is expressed only through relative transformations within the graph. As all nodes are dynamic, this eliminates the need for recalibration if a global reference changes or becomes unavailable, contributing to enhancing robustness and operational flexibility.

\subsection{Tracking by Querying Edges in DSG}
The proposed framework optimizes tracking accuracy using a global pose optimization over the DSG. As opposed to conventional sensor fusion problems \cite{ren2014}, both the tracking devices and the tracked entities in DSG are allowed to move at any time. Here, we define three key transformations in $\mathbf{SE}(3)$:
\begin{itemize}
    \item $\mathbf{P}_j$: the pose of passive node (marker) $j$,
    \item $\mathbf{A}_i$: the pose of active node (sensor) $i$,
    \item $\mathbf{T}_{ij}$: the relative pose of marker $j$ measured by sensor $i$.
\end{itemize}

The measurement is given by:
\[
\mathbf{T}_{ij} = \mathbf{A}_i^{-1}\mathbf{P}_j,
\]
where $\mathbf{A}_i$ and $\mathbf{P}_j$ represent the rigid body transformations described in the world reference. Practically, obtaining $\mathbf{A}_i$ involves complex setup and calibration because it depends on external localization. The DSG bypasses this challenge by focusing solely on the relative poses of the passive nodes, which can be calculated by: 

\[
\mathbf{P}_{j_1} ^{-1} \mathbf{P}_{j_2} = (\mathbf{A}_i \mathbf{T}_{ij_1})^{-1} \mathbf{A}_i \mathbf{T}_{ij_2} \implies \mathbf{P}_{j_1} ^{-1} \mathbf{P}_{j_2} = \mathbf{T}_{ij_1}^{-1} \mathbf{T}_{ij_2}
\]

The key observation here is that when computing relative poses between markers, the sensor's pose has no contribution in the final computation of intra-layer relationships. This means that the relative transformation between markers, which is our variable of interest, is invariant to the absolute sensor poses. Therefore, the detailed calibration and localization of $\mathbf{A}_i$ are not required for an accurate estimation of the relative marker poses.

\subsection{Measurement Optimization}

Given a set of valid inter-layer measurements, we refine the poses of all active nodes and passive nodes by solving a PGO problem over the DSG. For each $(i,j)$ in DSG, the measurement model is
\begin{equation}
\mathbf{T}_{ij} = \omega \mathbf{A}_i^{-1}\mathbf{P}_j,
\end{equation}
where $\mathbf{A}_i$ and $\mathbf{P}_j$ are unknown poses to be estimated, and $\omega$ is the binary status function describing tracking validity. Given current state estimates, the predicted measurement is:
\begin{equation}
\hat{\mathbf{T}}_{ij} \triangleq \mathbf{A}_i^{-1}\mathbf{P}_j.
\end{equation}

We define a residual on $\mathbf{SE}(3)$ and map it to $\mathfrak{se}(3)$ using the logarithm map:
\begin{equation}
\mathbf{r}_{ij} \triangleq \operatorname{Log}\!\left(\mathbf{T}_{ij}^{-1}\hat{\mathbf{T}}_{ij}\right)
= \operatorname{Log}\!\left(\mathbf{T}_{ij}^{-1}\mathbf{A}_i^{-1}\mathbf{P}_j\right)\in\mathbb{R}^6.
\end{equation}

The overall cost function is formulated as the sum of squared residuals weighted by an information matrix $\boldsymbol{\Omega}_{ij}$:
\begin{equation}
C\bigl(\{\mathbf{A}_i\},\{\mathbf{P}_j\}\bigr)
= \sum_{(i,j)\in V} \left \| \mathbf{r}_{ij}^{\mathsf{T}}\boldsymbol{\Omega}_{ij}\mathbf{r}_{ij} \right \|_2,
\label{eq:cost}
\end{equation}
where $\boldsymbol{\Omega}_{ij}$ represents the validity and confidence of the measurement ($\boldsymbol{\Omega}_{ij}=\mathbf{0}_{6\times6}$ if lose track). This cost function is equivalent to a least-square Maximum A Posteriori (MAP) estimation cost on $\mathbf{SE}(3)$. And we can therefore optimize~\eqref{eq:cost} by iteratively linearizing around current estimates. Let $\bar{\mathbf{A}}_i$ and $\bar{\mathbf{P}}_j$ be the current estimates, and apply perturbations in $\mathfrak{se}(3)$, then:
\begin{equation}
\mathbf{A}_i = \bar{\mathbf{A}}_i \exp(\delta\mathbf{A}_i),\qquad
\mathbf{P}_j = \bar{\mathbf{P}}_j \exp(\delta\mathbf{P}_j),
\end{equation}
where $\delta\mathbf{A}_i,\delta\mathbf{P}_j\in\mathbb{R}^6$ are small increments.

To resolve the global gauge freedom, we anchor one reference active node by fixing
\begin{equation}
\mathbf{A}_{i_0} = \mathbf{I}.
\end{equation}

Stacking all increments into a single vector $\delta$, the residuals are first-order approximated as
\begin{equation}
\mathbf{r}(\delta)\approx \mathbf{r}_0 + \mathbf{J}\delta,
\end{equation}
where $\mathbf{r}_0$ is evaluated at $(\{\bar{\mathbf{A}}_i\},\{\bar{\mathbf{P}}_j\})$, and $\mathbf{J}$ is the Jacobian of the stacked residual with respect to $\delta$.

We solve for $\delta$ using the Gauss-Newton update:
\begin{equation}
\mathbf{J}^{\mathsf{T}}\mathbf{J} \delta
= -\mathbf{J}^{\mathsf{T}}\mathbf{r}_0,
\label{eq:lm}
\end{equation}
 The updated poses are obtained by applying the computed increments, and~\eqref{eq:lm} is iterated until the convergence of \eqref{eq:cost} is reached.

\subsection{Scene Completion Under Occlusion}
Scene completion ensures continuity in the digital augmentation of physical spaces. Using the DSG and pose optimization, our framework can effectively compute the missing transformations between nodes when tracking is momentarily lost. Based on the third assumption, active nodes require shared visual access to at least one target with other trackers. \cref{fig:DSG}a illustrates the integration of the kinematic chain within the AR experience. As depicted in \cref{fig:DSG}b, in instances where a passive entity is no longer visible to an AR device, likely due to occlusion or movement out of the field of view, the DSG will establish a kinematic chain. This chain acts as a bridge, connecting the unobserved entity to the headset through other visible entities.

\begin{algorithm}
\caption{Kinematics Completion}
\label{alg:ar}
\begin{algorithmic}
\REQUIRE Dynamic Scene Graph $G(\{ n_p , n_a \})$
\REQUIRE Empty path set $P$

\FOR{All node paris in $\{ n_a, n_p \}$}
\STATE $P$ $\gets$ $\{\}$
\IF { isTracked($n_a$, $n_p$) }
\STATE $P$ = getEdge($n_a$, $n_p$)

\ELSE 
\STATE $P$ = DepthFirstSearch ($G$, $n_p$, $n_a$)
\ENDIF

\FOR{$ p = P$.pop($0$)}
\STATE    applyTransform ($H$, $p$)
\ENDFOR
\ENDFOR

\end{algorithmic}
\end{algorithm}

\cref{alg:ar} is applied to recover the missing transformations. When direct tracking becomes infeasible, the AR headset initiates a depth-first graph search algorithm to determine the most suitable path between the tracker and the obscured target. The computed path $P$ is a sequence of edges. By consecutively applying the previously computed optimal relative transformations, the position of the blocked target is derived. This newly acquired pose information is then utilized to render the digital model of the target accurately within the AR interface, reconstituting the integrity of the virtual overlay and ensuring a complete augmentation of the surgical scene.

\subsection{User Interface and Software Architecture}

As depicted in \cref{fig:Overview}b, the proposed system contains two components. The software is hosted on the central server and communicates with every sensor through TCP/IP. The software features parallel processing on the backend by separating the communications for each tracker into a standalone thread. On the front end of the software, we develop a PyQT-based Graphical User Interface. On the backend, the software maintains the DSG and dispatches the optimized poses to every sensor. 

To ensure adaptability in realistic environments, the software provides a unified sensor interface that abstracts away device-specific implementations. Through this interface, tracking devices can be integrated into the DSG independently of their underlying sensor models, communication protocols, or measurement modalities. As a result, the system supports real-time addition and removal of trackers during runtime without requiring system reinitialization, substantially improving operational flexibility and fault tolerance in the presence of sensor dropout or tracking loss due to occlusion.

\begin{figure}[thb]
\centering
\includegraphics[width=0.9\linewidth]{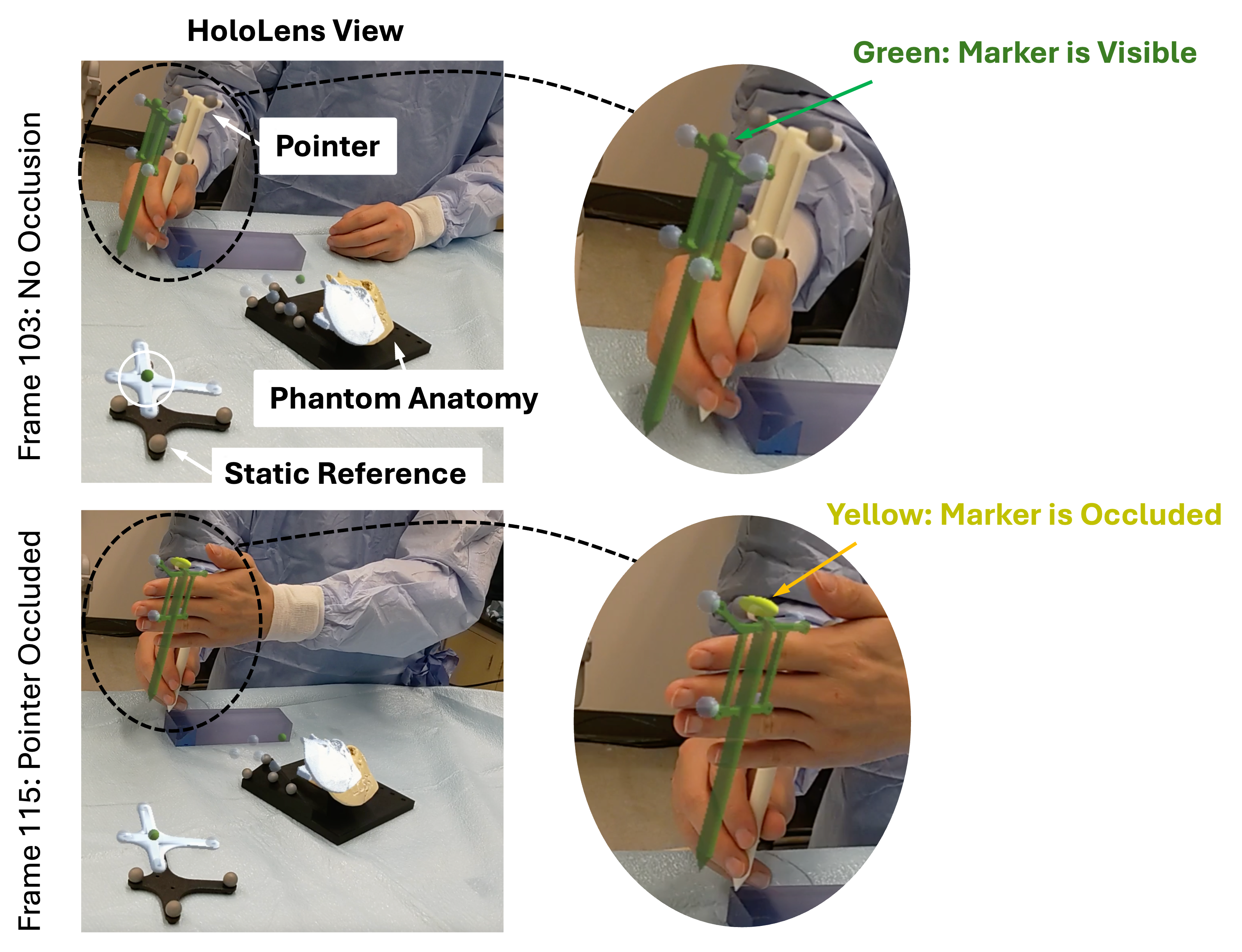}
  \caption{%
  	Example illustrating how our framework compensates for tracking loss. \textbf{Top}: At frame 103, all three objects are directly visible to the HoloLens, and a green indicator is displayed next to each rendered mesh to denote reliable tracking. \textbf{Bottom}: After the pointer is occluded by a hand at frame 115, the green indicator transitions into a yellow uncertainty ellipsoid, which represents the estimated pose uncertainty of the restored object based on the optimized dynamic scene graph. Note that the apparent misalignment between the virtual and physical objects is caused by the mixed reality capture functionality of the HoloLens, and not due to tracking errors.
  }
  \label{fig:losstrack}
\end{figure}

To convey tracking reliability to the user, the AR-HMD additionally visualizes pose uncertainty by rendering an ellipsoid whose principal axes correspond to the estimated confidence of the pose in real time, as illustrated in \cref{fig:losstrack}. The direction of the axes is aligned with the local reference frame of each marker. When a direct line of sight (LoS) between the AR-HMD and the tracked marker is available, the ellipsoid is rendered in green, indicating that the pose is directly observed by the headset. In contrast, when LoS is lost, the ellipsoid transitions to yellow, informing the user that the pose is inferred through optimization using alternative kinematic constraints within the DSG.

The magnitude of uncertainty is encoded in the length of the ellipsoid’s principal axes. This magnitude is proportional to the diagonal entries of the inverse of the corresponding Hessian matrix, obtained by:
\[
H_j^{-1} = (J_j^\top J_j)^{-1},
\]
where $J_j$ denotes the Jacobian associated with passive node $j$ after convergence. Smaller diagonal values of $H_j^{-1}$ indicate stronger local observability and lower uncertainty, resulting in a more compact ellipsoid. On the contrary, weaker constraints yield longer axes. This visualization provides users with an explicit representation of tracking confidence, enhancing situational awareness during AR-assisted procedures, particularly in the presence of intermittent tracking loss.

\section{Experiments and Results}

\subsection{Simulation Evaluation} 

To assess the effectiveness of our tracking optimization approach in simulation, we use synthetic data consisting of two targets and two sensors. At each time step, we compare the relative transformation between the two targets as estimated independently from measurements provided by the two sensors.

The two targets belong to the passive layer of the DSG, while the sensors are modeled as active nodes. Both sensors and targets undergo motion with randomized linear and angular accelerations. Over a simulated duration of 500 seconds, the targets traverse a distance of $82.7~\mathrm{m}$, during which measurements are randomly occluded with probability $p_{\text{block}} = 0.1$. To quantify tracking performance, we employ the absolute trajectory error (ATE) to measure global pose accuracy and the relative trajectory error (RTE) to capture local drift accumulation, both reported as root mean square error (RMSE).

Let ${\mathbf{T}i^{\text{gt}}}, i \in (1,N)$ denote the ground-truth poses of a target and ${\mathbf{T}i^{\text{est}}}, i \in (1,N)$ the corresponding estimated poses. Following the Umeyama alignment procedure~\cite{umeyama_least-squares_1991}, a similarity transformation $\mathbf{S} \in SE(3)$ is computed to optimally align the estimated trajectory to the ground truth. The ATE is then defined as
\begin{equation}
\mathrm{ATE}=
\sqrt{
\frac{1}{N}
\sum_{i=1}^{N}
\left|
(\mathbf{T}_i^{\text{gt}})^{-1}
\mathbf{S}
\mathbf{T}_i^{\text{est}}
\right|^2
}
\end{equation}

To evaluate relative drift, we compute the RTE over a fixed temporal interval $\Delta$. For each valid index $i$, the relative pose error is defined as

\begin{equation}
\mathbf{E}_i^{\text{rel}} = 
\left(
(\mathbf{T}i^{\text{gt}})^{-1}
\mathbf{T}_{i+\Delta}^{\text{gt}}
\right)^{-1}
\left(
(\mathbf{T}i^{\text{est}})^{-1}
\mathbf{T}_{i+\Delta}^{\text{est}}
\right),
\end{equation}

RTE is computed as the average of $\mathbf{E}^{\text{rel}}$ within $\Delta$. Using our framework to integrate measurements from both sensors, we obtain an ATE of $9.12~\mathrm{mm} / 3.65^\circ$ and an RTE of $4.79~\mathrm{mm} / 5.17^\circ$. In contrast, using measurements from a single sensor results in a substantially higher ATE of $24.61~\mathrm{mm} / 9.86^\circ$ and an RTE of $33.20~\mathrm{mm} / 13.95^\circ$. These results demonstrate that multi-sensor fusion within the DSG significantly mitigates the impact of intermittent tracking loss, improves pose consistency, and contributes to suppressing drift accumulation.

Simulation is further used to compare our framework against conventional calibration-based sensor fusion approaches~\cite{naheemOptical2022, heInertial2015}. Because such methods depend on restrictive assumptions, like using static sensors with known extrinsic calibration, the simulation setup is intentionally simplified to satisfy their operational requirements. In particular, sensors are held fixed while target trajectories are replayed, enabling a fair and controlled comparison under conditions in which calibration-based fusion is applicable.

Under these conditions, we evaluate performance using ATE, the standard deviation of ATE, the top 5\% largest translational errors, and the ratio of time when tracking is lost against the whole duration of the experiment. As summarized in \cref{tab:simexp}, the proposed framework achieves comparable accuracy to calibration-based fusion when sensors remain static, demonstrating that DSG-based optimization preserves performance in regimes where classical methods are valid. Importantly, unlike calibration-dependent approaches, our framework does not rely on fixed sensor configurations and remains fully functional when both sensors and targets are dynamic. These results show that the proposed method not only matches existing techniques but also generalizes to more complex and realistic scenarios that cannot be handled by conventional fusion approaches.

\begin{table}[ht]
\centering
\caption{Simulation Trajectory Errors Comparison between Calibration-based Sensor Fusion and DSG$+$PGO}
\label{tab:simexp}
\begin{adjustbox}{width=0.47\textwidth}
\begin{tabular}{lcccc}
\toprule
 Target ID & Approach & ATE Trans $\pm$ SD [mm]  & top 5\% RPE Trans [mm] & Loss Track Ratio [\%]  \\
\midrule
\multirow{2}{*}{Target 1} 
& Fusion & 5.49 $\pm$ 3.12 & 2.33 & 0.10 \\
& Ours & 5.20 $\pm$ 3.68 & 2.46 & 0.04 \\
\midrule
\multirow{2}{*}{Target 2} 
& Fusion & 6.01 $\pm$ 4.74 & 3.25 & 0.10 \\
& Ours   & 5.88 $\pm$ 4.26 & 3.97 & 0.03 \\
\bottomrule
\end{tabular}
\end{adjustbox}
\end{table}

\subsection{Real-world Experiment}
\subsubsection{Trajectory Error}
\begin{figure}[htb]
  \centering 
  \includegraphics[width=\columnwidth]{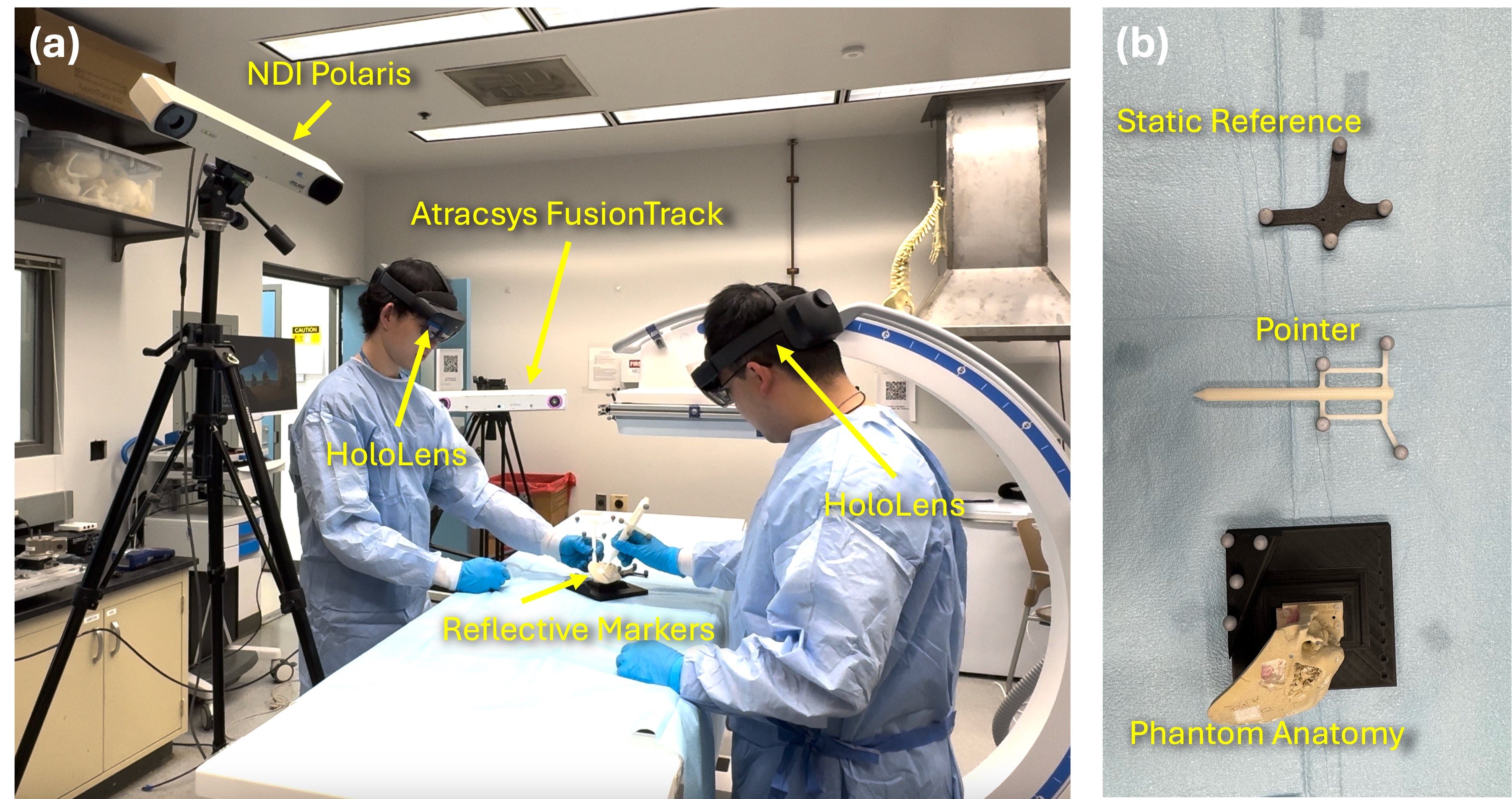}
  \caption{%
  	(a) Experiment setup for a real-world quantitative experiment. Two operators were wearing HoloLens 2 for tracking and real-time rendering of the tracked markers. An NDI Polaris optical tracker and an Atracsys FusionTrack optical tracker were placed next to the operating table with their view point set to cover the table. The operators used pointers to mimic a drilling procedure on a phantom anatomy. (b) Three types of markers were used in the experiment. Each marker is equipped with four reflective spheres.
  }
   \label{fig:realexp}
\end{figure}

 The central server is running on a workstation with an Intel Core i7-9750 CPU and 16GB of RAM. We integrated three distinct types of tracking systems, namely NDI Polaris, Atracsys TrackFusion, and Microsoft HoloLens 2. These tracking systems were used to track three unique instruments, each outfitted with four retro-reflective spheres, and were placed such that they share an FoV on the operating table. Among these systems, Atracsys FusionTrack was designated as the ground-truth reference. This device provides the highest nominal accuracy, achieving sub-millimeter precision, and it is actively used for surgical navigation purposes \cite{shuTwinS2023, naheemOptical2022}. Two operators were wearing HoloLens 2 to monitor the rendering of the tracked entities. The setup of the experiments is illustrated in \cref{fig:realexp}. The data from the moving target can have larger noise than that from a static target. In light of such a limitation, we placed one of the markers relatively affixed to the operating table. Another rigid body marker is attached to a 3D-printed pointer and held by the operator. 

The Microsoft HoloLens 2 leverages its Articulated Hand Tracking (AHAT) camera system to detect the spheres and reference them against predefined rigid body templates using STTAR \cite{martin-gomez2023}. The tracked data were then communicated to the server via the TCP/IP protocol. Concurrently, the NDI Polaris system and Atracsys FusionTrack 500 system utilized dual cameras to triangulate marker positions from calibrated multiangle scene views, with data relayed through an Ethernet connection using the device's native driver.

We calculated the pointer tip trajectory \textit{w.r.t.} static reference measured from both NDI Polaris and HoloLens 2 as a baseline of single device tracking. The proposed framework fuses the measurements to get the optimized relative trajectory of the pointer tip. To assess the tracking error of both the HoloLens 2 and NDI camera systems, we first mitigate the latency inherent in TCP/IP transmission. Following this, we employ the Umeyama algorithm to precisely register the tracked trajectories \textit{w.r.t.} the NDI coordinate system. Subsequently, we compute the ATE at each time step to evaluate global consistency, while the RPE is computed to gauge drift for individual tracks.

\begin{figure}[t]

  \centering 
  \includegraphics[width=1.0\linewidth]{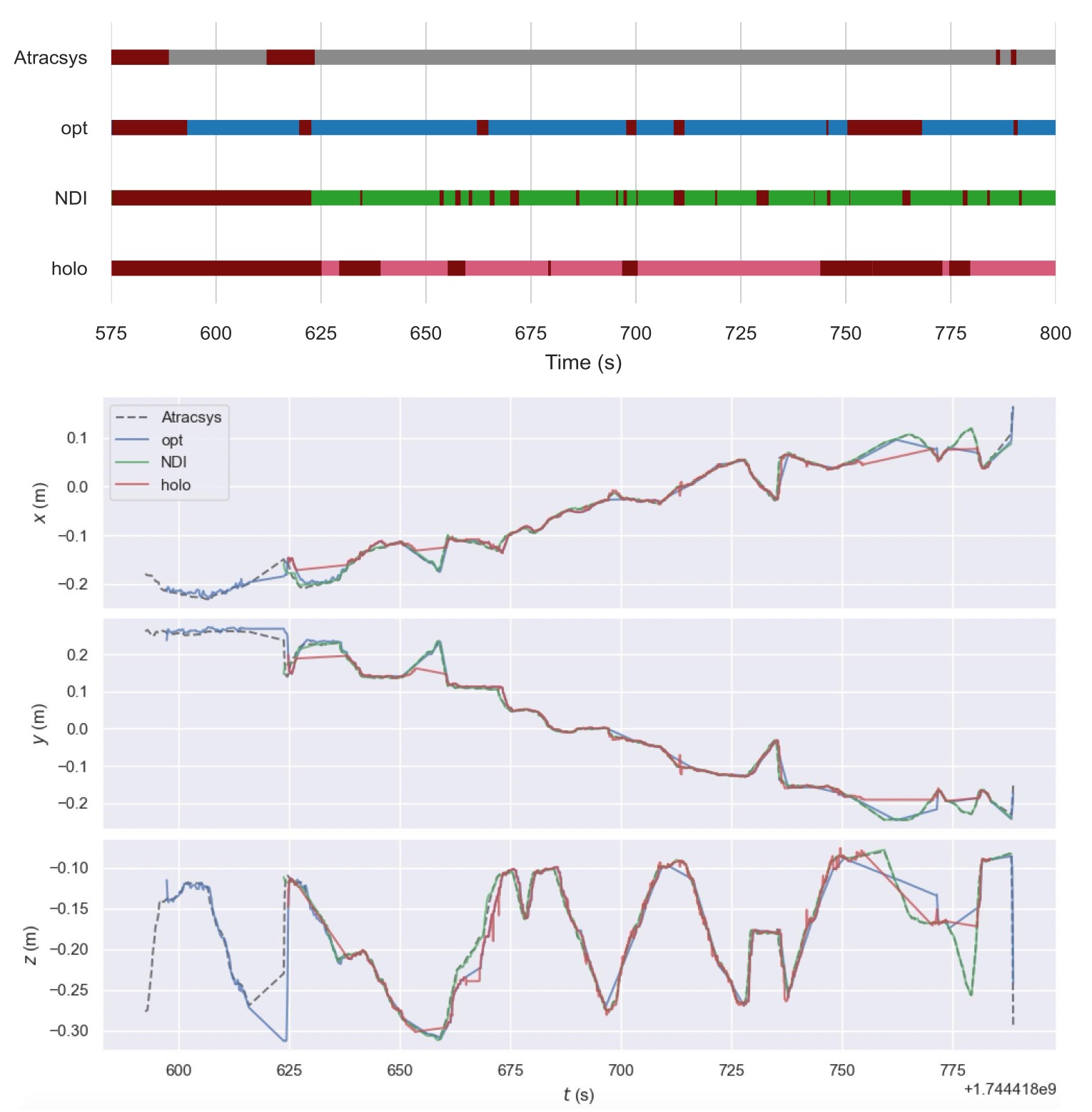}
  \caption{A segment of tracking for the ground truth, the HoloLens tracking, the NDI tracking, and our framework. Top panel shows the tracking status (tracked or lose track), and the bottom panel shows the trajectories decomposed into three axes.}
    \label{fig:proj_traj}
\end{figure}

\begin{table}[ht]
\centering
\caption{Real World Experiment Trajectory Errors}
\label{tab:realexp}
\begin{adjustbox}{width=0.48\textwidth}
\begin{tabular}{lccccc}
\toprule
Exp. & Source & ATE Trans [m] & RPE Trans [m] & Loss Track Ratio [\%] & Total Length [m] \\
\midrule
\multirow{3}{*}{A} 
 & HoloLens & 0.27 & 0.08 & 30.6 & \\
 & NDI      & 0.77 & 0.33 & 56.5 & 3.49 \\
 & Ours     & \textbf{0.03} & \textbf{0.01} & \textbf{19.4} & \\
\midrule
\multirow{3}{*}{B} 
 & HoloLens & 0.24 & 0.07 & 52.7 & \\
 & NDI      & 0.99 & 0.38 & 79.9 & 3.56 \\
 & Ours     & \textbf{0.09} & \textbf{0.07} & \textbf{33.2} & \\
\midrule
\multirow{3}{*}{C} 
 & HoloLens & 0.45 & 0.21 & 39.9 & \\
 & NDI      & 0.44 & 0.10 & 15.9 & 3.87 \\
 & Ours     & \textbf{0.05} & \textbf{0.03} & \textbf{13.2} & \\
\bottomrule
\end{tabular}
\end{adjustbox}
\end{table}

\begin{figure*}[ht]

  \centering 
  \includegraphics[width=0.9\linewidth]{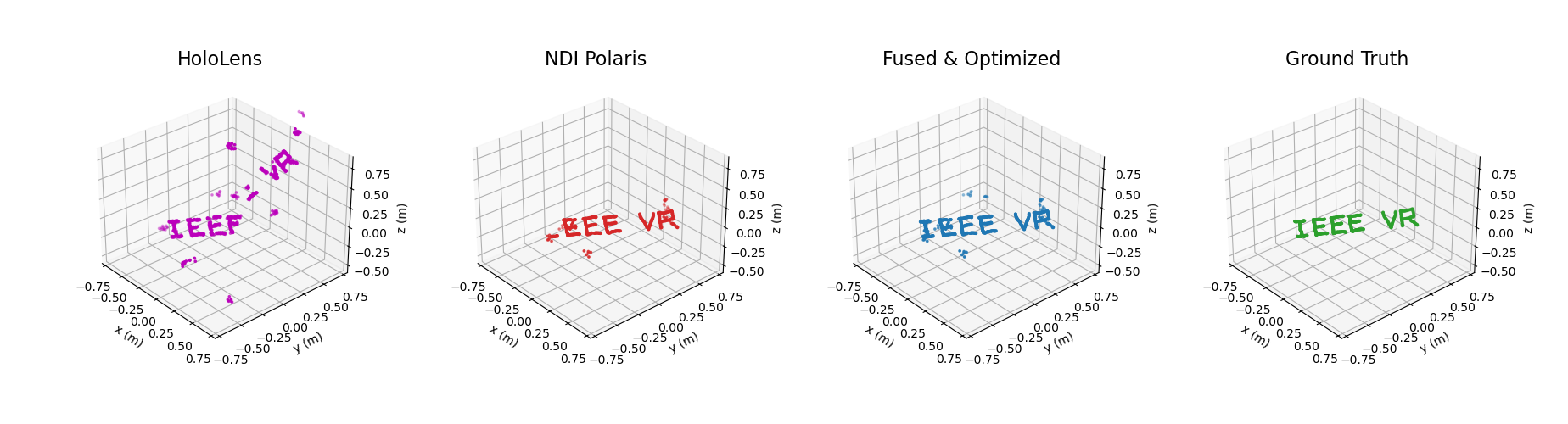}
  \caption{%
  	Example measured trajectory recorded when drawing an "IEEE VR" text with a pointer. HoloLens, although it covers a larger area of the text, has the most outliers, and its letter "VR" drifted from the main trajectory, indicating a wrong detection of the markers. As opposed to HoloLens, NDI Polaris has better tracking consistency, yet due to its limited FoV, the letter "I" is untrackable. Our proposed fusion can leverage the advantages of both sensors and outperforms each of the single sensors.
  }
    \label{fig:VR}
\end{figure*}

The tracking results of the three experimental scenarios are compared in \autoref{tab:realexp}. \cref{fig:proj_traj} provides a segment of the trajectory. We reported errors for each individual tracker (NDI and HoloLens 2) and our fusion method and compared them with the Atracsys trajectories, which are considered the ground truth. It is important to note that we track the relative poses between two markers using different sensors. Similar to previous analyses, we calculate both translational and rotational errors relative to optical tracking. 

In situations where sensors or objects are in motion, occasional tracking loss would occur on the sensor used for ground-truth. We disregarded these poses as they would introduce unnecessary bias during evaluation. The segment in \cref{fig:proj_traj} shows that although NDI achieves accuracy comparable to the ground-truth system during valid observations, it experiences a higher frequency of tracking loss than the other devices, primarily due to its limited field of view and the intentionally introduced occlusions. Overall, our fusion tracking demonstrates improved robustness and comparable accuracy compared to both the HoloLens and NDI in all experiments, yielding more stable and reliable trajectories than using either system individually.

\subsubsection{Scene Completion}
The proposed system could compensate for tracking loss in most cases. Compared to any single device system, our fusion method dynamically adjusts confidence levels to sensor readings, ensuring robust and consistent tracking with relatively high precision. To verify the improvement in completeness, we use the pointer to draw characters on the operation table. An example of  "IEEE VR" is shown in \cref{fig:VR}. Although the waypoints on the trajectory measured only with HoloLens are pretty dense, it has significant drifting and an obvious number of outliers; on the other hand, the NDI's trajectory is consistent, but completely loses tracking when writing the letter "I". With our scene graph, we can effectively fuse these results to leverage the advantage of both sensors. Compared to the ground truth trajectory, the optimized trajectory covers most regions of the letters and maintains a consistent spatial relationship.   

In one mock-up experiment, we move the pointer around the operating table and try to randomly block it from the sensors. As shown in \cref{fig:losstrack}, all three objects are visible to HoloLens at frame 103, with a green dot floating next to the rendered mesh of the tracked objects, indicating that direct tracking from HoloLens to the pointer is established. Then, we intentionally block the pointer at frame 115 to trigger a scene completion that utilizes other external sensors to obtain the position of the pointer in HoloLens's viewpoint. The green dot will transit into a yellow ellipsoid, the axis lengths of which indicate the uncertainty of the calculated pointer position at the x, y, and z directions.

\section{Discussion}

The inherent complexity of surgical environments demands reliable and precise guidance during AR-assisted procedures. Our framework addresses this challenge by integrating pose data from multiple devices into a dynamic scene graph and optimizing it globally. This approach enables robust alignment of digital models with real-world targets, effectively mitigating occlusion and limited fields of view. As a result, the system enhances the stability and precision of AR guidance, supporting safe and efficient surgical performance.

A central strength of the framework is its device-agnostic design, which allows plug-and-play integration with diverse tracking technologies. Although our current implementation relies on marker-based tracking, the system architecture readily accommodates more advanced modalities, including markerless pose estimation, as well as additional sensing technologies. This adaptability enables intraoperative flexibility, such as dynamically adding or removing trackers without disrupting workflow. These capabilities provide opportunities to empower surgeons and assist them in making more informed decisions during complex procedures. Furthermore, while our experiments focus on surgical settings, the general design allows straightforward application to other specialized domains that require accurate and flexible tracking.

Despite these strengths, several limitations remain. First, the current implementation is primarily marker-based, which may restrict applicability in scenarios where attaching fiducials is impractical or undesirable. Extending the framework to fully exploit feature-based or frameless tracking would further increase its clinical adoption. Second, computational performance remains a bottleneck. At present, global optimization requires $0.14 \pm 0.09$ seconds per graph update. While acceptable in small- to medium-scale settings, larger numbers of data sources may increase latency, risking frame drops in rendering on end devices such as AR-HMDs. Third, although the system is designed to be device-agnostic, a broader combination of commonly used tracking sensors, such as IMUs and EM systems, is not evaluated and discussed. These limitations highlight opportunities for future work, particularly GPU acceleration and integrating more sensor modalities.

In addition, we acknowledge that the real-world experiments presented in this work may not fully represent the complex variability and unpredictability of real surgical environments. This experimental design was deliberately made to validate the core contribution of the proposed framework, namely, its ability to maintain globally consistent tracking through multi-view optimization without confounding factors introduced by procedure-specific workflow variability or uncontrolled clinical constraints. Comprehensive evaluations under highly complex surgical conditions will be the focus of future studies, alongside larger-scale experiments and surgery-specific validation.

Looking beyond surgical use, the proposed framework also shows promise for broader AR applications. In collaborative AR settings, the ability to integrate multiple heterogeneous sensors into a unified dynamic scene graph can facilitate consistent multi-user visualization. Similarly, in industrial AR applications such as assembly, inspection, or maintenance, the system's adaptability and resilience to occlusion can improve both efficiency and safety. These potential extensions underline the versatility of the framework and position it as a generalizable solution for multi-sensor, real-time AR tracking across domains.

\section{Conclusion}

 This work presents a novel tracking framework that integrates multiple data sources into a dynamic scene graph and optimizes them in real-time through pose graph optimization. By explicitly modeling sensor uncertainties and removing reliance on a fixed global reference, the framework achieves resilient and device-agnostic tracking that provides consistent AR visualization even under occlusion. Experimental results demonstrated improved tracking stability and reduced loss ratios compared to single-device baselines, particularly in scenarios with frequent line-of-sight interruptions.

\section*{Supplemental Materials}
 To assist the reproduction of this work, code is available at \href{https://github.com/jmz3/PoseHub.git}{GitHub}.

\acknowledgments{
This paper is partially supported by NIH under Grant No. R01DC018815 and R01AR080315.}
\bibliographystyle{abbrv-doi}

\bibliography{template}
\end{document}